\documentclass[twoside,reqno]{HERON}
\usepackage{epsfig,cite,colordvi}
\usepackage{graphicx}
\usepackage{url}
\usepackage{amssymb,amsmath,amscd,epsf}
\usepackage{times}
\usepackage{makeidx}

\newcommand{\bea}{\begin{eqnarray}}
\newcommand{\eea}{\end{eqnarray}}
\newcommand{\be}{\begin{equation}}
\newcommand{\ee}{\end{equation}}
\newcommand{\np}{{\bf p}}

\newcommand{\nh}{{\bf h}}
\newcommand{\nk}{{\bf k}}

\newcommand{\nq}{{\bf q}}

\newcommand{\sumint}{\sum\kern -3.5ex \int\kern 1.0ex}

\begin{document}

\title{The role of meson exchange currents in charged current (anti)neutrino-nucleus scattering}

\runningheads{The role of meson exchange currents in neutrino-nucleus scattering}{M.B. Barbaro, \etal}

\begin{start}

\author{M.B. Barbaro}{1,2}, \coauthor{J.E. Amaro}{3}, \coauthor{J.A. Caballero}{4}, \coauthor{A. De Pace}{2}, \coauthor{T.W. Donnelly}{5}, \coauthor{G.D. Megias}{4}, \coauthor{I. Ruiz Simo}{3}

\index{Barbaro, M.B.}
\index{Amaro, J.E.}
\index{Caballero, J.A.}
\index{De Pace, A.}
\index{Donnelly, T.W.}
\index{Megias, G.D.}
\index{Ruiz Simo, I.}

\address{Dipartimento di Fisica, Universit\`a di Torino, 10125 Torino, Italy}{1}

\address{INFN, Sezione di Torino, 10125 Torino, Italy}{2}

\address{Departamento de F\'isica At\'omica, Molecular y Nuclear and Instituto de F\'isica Te\'orica y Computacional Carlos I, Universidad de Granada, 18071 Granada, Spain}{3}

\address{Departamento de F\'isica At\'omica, Molecular y Nuclear, Universidad de Sevilla, 41080 Sevilla, Spain}{4}

\address{Center for Theoretical Physics, Laboratory for Nuclear Science and Department of Physics, Massachusetts Institute of Technology, Cambridge, MA 02139, USA}{5}

\begin{Abstract}

We present our recent progress in the description of neutrino-nucleus interaction in the GeV region, of interest for ongoing and future oscillation experiments. In particular, we discuss the weak excitation of two-particle-two-hole states induced by meson exchange currents in a fully relativistic framework. 
We compare the results of our model with recent measurements of neutrino scattering cross sections, showing the crucial role played by two-nucleon knockout in the interpretation of the data.
\end{Abstract}
\end{start}

\section{Introduction}

The accurate understanding of neutrino-nucleus scattering in the GeV region is a challenging problem of many-body nuclear physics and a necessary input for particle physics studies related to neutrinos.
The modeling of this reactionis indeed the largest and most complicated source of uncertainty in present neutrino long-baseline experiments (T2K, NOvA, MicroBooNE) and it will be a crucial limitation for the sensitivity of next generation neutrino oscillation experiments (Hyper-Kamiokande, DUNE).
These experiments, as other ones already completed (MiniBooNE, SciBooNE, NOMAD, ArgoNeuT), aim at precision measurements of the oscillation parameters; most importantly, they search for signals of leptonic CP violation, which may help explaining the matter-antimatter asymmetry of the Universe.
Being the detectors made of large volumes of complex nuclei (typically carbon, oxygen and argon), good control of neutrino-nucleus interactions is required in order to extract significant information on the neutrino physics. On the other hand, the same experiments can provide interesting informations on the nuclear dynamics, complementary to the ones obtained using electromagnetic and hadronic probes. This has motivated a very intense activity of nuclear theorists on the subject over the last few years. Comprehensive reports on the state of the art of the field can be found in several review articles \cite{GGZReview, FZReview, AHNReview, UMReview}.

One of the main difficulties in the analysis of the above mentioned experiments is the lack of precise knowledge of the incoming neutrino energy $E_\nu$, which is broadly distributed around an average value. 
As a consequence, $E_\nu$ can only be reconstructed from the outgoing particle(s) kinematics; this procedure is strongly dependent upon the model adopted to describe the nucleus in its initial and final states, on the treatment of final-state interactions (FSI) between the outgoing nucleon(s) and the residual nucleus, as well as on the assumptions about the different mechanisms contributing to the observed cross section.
The theoretical description of these effects is a complicated many-body problem.
A wide variety of models has been proposed to describe CCQE (Charged-Current-Quasi-Elastic) neutrino and antineutrino cross sections, identified experimentally by the absence of pions in the final state. 
These models rely on quite different hypotheses and approximations and utilize diverse theoretical frameworks: 
Relativistic Mean Field Theory \cite{Gonzalez},
Random-Phase-Approximation \cite{Martini,Nieves,Ghent}, 
Relativistic Green Function \cite{Giusti},
Spectral Function Formalism\cite{Benhar},
Green Function Monte Carlo \cite{Schiavilla}, 
Coherent Density Fluctuation Model \cite{Antonov},
Transport Theory \cite{Mosel}.
Despite these differences, there is now a general consensus on the important role played by two-particle-two-hole (2p2h) excitations in the interpretation of the CCQE data. Such excitations are induced by two-body currents, hence they go beyond the usual impulse approximation scheme, in which the probe interacts with one single nucleon. Nevertheless, due to the above mentioned definition of CCQE events in neutrino experiments (no pions in the final state), they also contribute to the observed cross section.

Though the present precision of neutrino experimental data is generally too low to discriminate between different models,  
valuable informations on the reliability of the latter - in a given kinematical regime - are provided by the large amount of high quality electron scattering data. The validation against these data represents a mandatory (albeit not sufficient) test any model should pass before being applied to neutrino scattering reactions, where other sources of uncertainties are present -- {\it e.g.} the limited knowledge of the elementary neutrino-nucleon form factors. 
In order to fulfill this requirement, we have developed a nuclear model based on some general properties of electron scattering data and known as SuSA (Super-Scaling-Approach). The model has been first introduced in Ref. \cite{Amaro:2004bs} and subsequently refined by including effects from Relativistic Mean Field (RMF) theory and adding the contribution of meson-exchange-currents (MEC) in the 2p2h sector: the resulting model (called SuSAv2-MEC) has been recently validated versus electron scattering data in a wide range of kinematical conditions \cite{Megias:2016lke}.
The model is capable of describing different kinematical regions: quasielastic, 2p2h and inelastic. For the neutrino case the latter is presently limited to the $\Delta$-resonance excitation, while for electron scattering the full inelastic spectrum is considered.
Here we will focus mainly on the treatment of the 2p2h region, which occurs for transferred energies between the QE and $\Delta$ peaks. 

The present contribution is organized in the following way: in Section \ref{sec:formalism} we briefly define the formalism needed for studying CC neutrino and antineutrino scattering, with particular reference to the 2p2h contribution. In Section \ref{sec:results} we present some selected results and compare our predictions with experimental data. Finally, in Section \ref{sec:conclusions} we summarize our work and outline future developments.

\section{Formalism}
\label{sec:formalism}

In this Section we summarize the essential formalism for $(\nu_l,l^-)$ and
antineutrino $(\overline{\nu}_l,l^+)$ CC reactions in nuclei. 
Let  $K^\mu=(\epsilon,\nk)$, $K'{}^\mu=(\epsilon',\nk')$ be the incident and scattered lepton four-momenta, respectively, $Q^{\mu}=(\omega,\nq)=(K-K')^{\mu}$
the four-momentum transfer, $\omega$ and $\nq$ the energy and momentum transfer.
The $z$ direction is chosen along the vector $\nq$.
The double-differential cross section is
\begin{equation}
\frac{d\sigma}{d\Omega'd\epsilon'}
= \sigma_0 {\cal S_{\pm}},
\label{sigma}
\end{equation}
where $\Omega'$ is the scattering solid angle and $\sigma_0$ is a kinematical factor including the weak couplings (see Appendix).  
The nuclear structure function ${\cal S_{\pm}}$ is the linear combination of five response functions
\begin{equation}
{\cal S_{\pm}}
=
V_{CC} R^{CC}
+ 2 V_{CL} R^{CL}
+ V_{LL} R^{LL}
+ V_{T} R^{T}
\pm 2 V_{T'} R^{T'},
\label{Spm}
\end{equation}
where the sign of the last term is positive for neutrinos and negative
for antineutrinos.
The $V_K$'s, whose expressions are given in the Appendix, are kinematical factors depending on the lepton's kinematics. The five response functions
\begin{eqnarray}
R^{CC} &=& W^{00} \label{rcc} \\
R^{CL} &=& -\frac12\left(W^{03}+ W^{30}\right) \\
R^{LL} &=& W^{33}  \\
R^{T} &=& W^{11}+ W^{22} \\
R^{T'} &=& -\frac{i}{2}\left(W^{12}- W^{21}\right) \label{rtp}
\end{eqnarray}
embody the nuclear structure and dynamics and are given in terms of components of the hadronic tensor 
\begin{equation}
W^{\mu\nu} = \overline{\sum_i} \sum_f \langle f | {\hat J}^\mu(Q)| i \rangle^* 
\langle f | {\hat J}^\nu(Q)| i \rangle\, \delta(E_i+\omega-E_f) .
\end{equation}
 In the above expression ${\hat J}^\mu(Q)$ represents the nuclear many-body current operator, the initial and final nuclear states $|i\rangle$ and $|f\rangle$ are exact eigenstates of the nuclear Hamiltonian, with energies $E_i$ and $E_f$, respectively, and the symbol $\overline{\sum}_i$ means average over initial states.
This form is very general and includes all possible final states that can be reached through the action of the current operator ${\hat J}^\mu(Q)$ on the exact ground state. 
Thus the hadronic tensor can be expanded as the sum of one-particle one-hole (1p1h), two-particle two-hole (2p2h), plus additional channels:
\begin{equation}
W^{\mu\nu} = W^{\mu\nu}_{1p1h} + W^{\mu\nu}_{2p2h} + \cdots
\end{equation}
Here we focus on 2p2h excitations, within a model which will be shortly presented in the next subsection.

\subsection{2p2h excitations}
\label{sec:MEC}

In order to evaluate the 2p2h hadronic tensor $W^{\mu\nu}_{2p2h}$, we choose to describe the nuclear ground state as a relativistic Fermi gas (RFG), characterized by a Fermi momentum $k_F$. 
The main justification for this choice, which is undoubtedly too simple to encompass all aspects of nuclear dynamics, is that we are interested in a kinematical region where typical energies are of the order of or higher than the nucleon mass and therefore relativistic effects cannot be neglected. 
The RFG model is one of the few nuclear models in which Lorentz covariance can be maintained. 
 
\begin{figure}[h]
\begin{center}
\includegraphics[scale=0.77,bb=160 200 454 686]{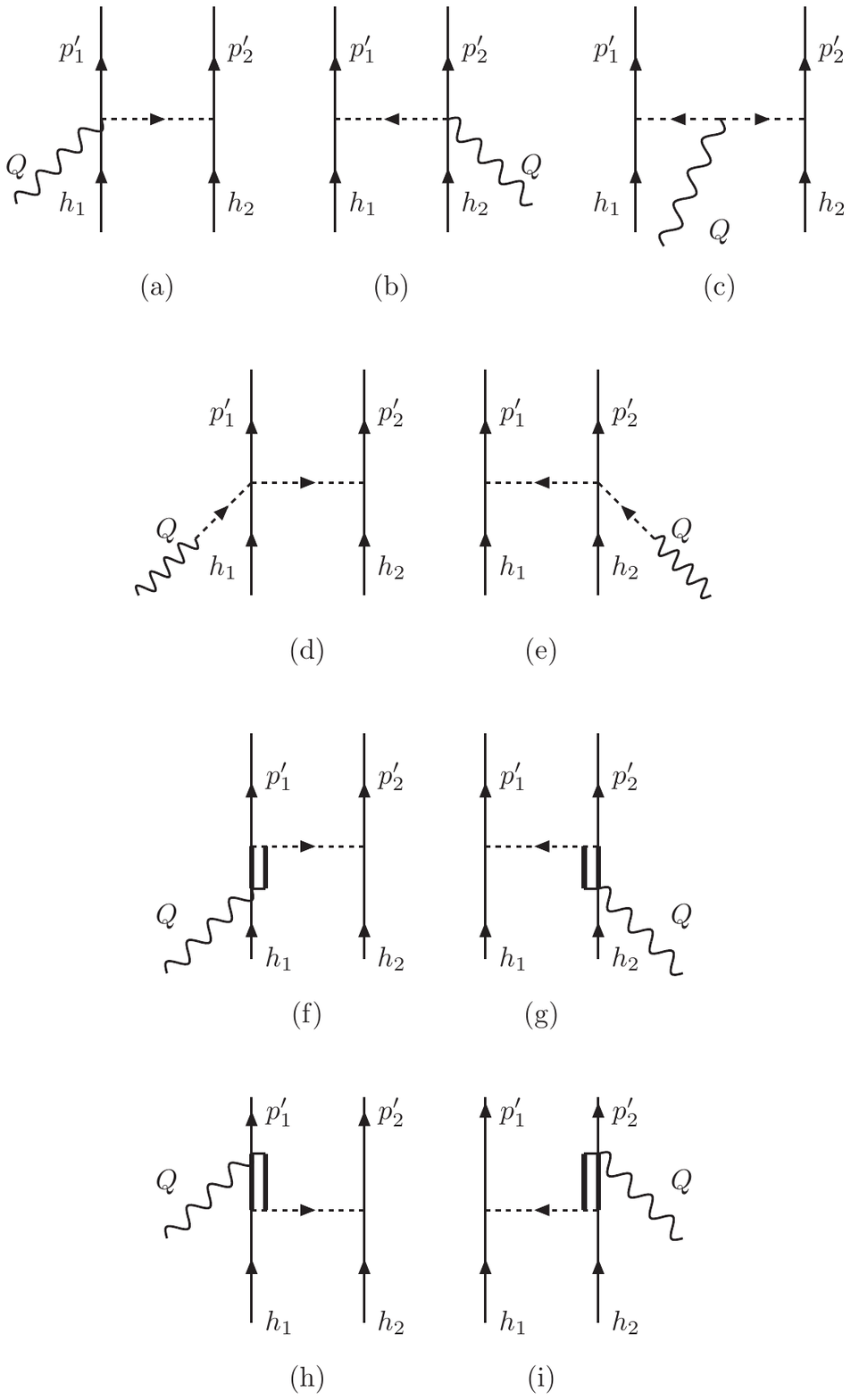}
\caption{ Feynman diagrams of the MEC considered in the present study,
  including the seagull (a,b), pion-in-flight (c), pion-pole (d,e),
  and $\Delta$ pole (f--i).  
}
\label{diagrams}
\end{center}
\end{figure}

In the RFG, the 2p2h channel corresponds to states with two nucleons of momenta $\np'_1$ and $\np'_2$
above the Fermi momentum, $p'_i>k_F$, and two hole states of momenta $\nh_1$ and $\nh_2$  below the Fermi momentum, $h_i<k_F$.
The  2p2h hadronic tensor is given by
\begin{eqnarray}
W^{\mu\nu}_{2p2h}
&=&
\frac{V}{(2\pi)^9}\int
d^3p'_1
d^3h_1
d^3h_2
\frac{M^4}{E_1E_2E'_1E'_2} \Theta(p'_1,p'_2,h_1,h_2)
\nonumber \\ 
&\times&
r^{\mu\nu}(\np'_1,\np'_2,\nh_1,\nh_2)
\delta(E'_1+E'_2-E_1-E_2-\omega) ,
\label{hadronic}
\end{eqnarray}
where $V$ is the volume of the system, $\bf p'_2= h_1+h_2+q-p'_1$  is fixed by  momentum conservation, $M$ is the nucleon mass, the energies $E_i$ and $E'_i$ are the on-shell energies of the holes and particles, and 
\begin{equation}
\Theta(p'_1,p'_2,h_1,h_2)
\equiv
\theta(p'_2-k_F)
\theta(p'_1-k_F)
\theta(k_F-h_1)
\theta(k_F-h_2).
\end{equation}
By exploiting the energy $\delta$-function and rotational symmetry the above expression can be reduced to a 7-dimensional integral, to be computed numerically.
The non-trivial part of the calculation is contained in the 
function $r^{\mu\nu}$, which represents the elementary hadronic tensor for the basic 2p2h transition, with given initial and final momenta, summed over spin ($s_i$, $s'_i$) and isospin  ($t_i$, $t'_i$) projections:
\begin{equation}
r^{\mu\nu}(\np'_1,\np'_2,\nh_1,\nh_2)=
\frac{1}{4}\sum_{s_1s_2s'_1s'_2}
\sum_{t_1t_2t'_1t'_2}
j^{\mu}(1',2',1,2)^*_A
j^{\nu}(1',2',1,2)_A .
\label{elementary}
\end{equation}
In the above, $j^{\mu}(1',2',1,2)_A$ is the antisymmetrized matrix element of the two-body weak MEC, containing vector and axial components. The present model is the extension to the axial sector of the calculation of Ref.~\cite{Arturo}.
We work at tree level and include only one-pion exchange. Then the MEC operator
can be written as the sum of four contributions, denoted as
seagull, pion-in-flight, pion-pole and Delta-pole and represented diagrammatically in Fig. \ref{diagrams}. The explicit expression are too long to be reported here and can be found in Ref. \cite{Simo:2016ikv}.

The sum over isospin in Eq.~(\ref{elementary}) combines all the possible charge channels
in the final state, corresponding to emission of $pp$, $nn$ and $pn$
pairs. CC neutrino scattering can induce two possible 2p2h transitions,
 $np \rightarrow pp$ and $nn \rightarrow np$, while in the case of 
antineutrino the allowed channels $np\rightarrow nn$ and $pp\rightarrow np$. 
In our formalism, it is possible to separate the contributions
of these charge states to the response functions, namely to know how many pairs of each kind ($pp$, $nn$ and $pn$) participate to this specific process.

\section{Results}
\label{sec:results}

Let us first present the predictions of our MEC model for the different response functions. We consider $\nu_\mu$-$^{12}$C scattering and typical kinematics of interest for current neutrino oscillation experiments.

 \begin{figure}[h]
   \vspace{-0.8cm}
   \includegraphics[scale=0.5, angle=0]{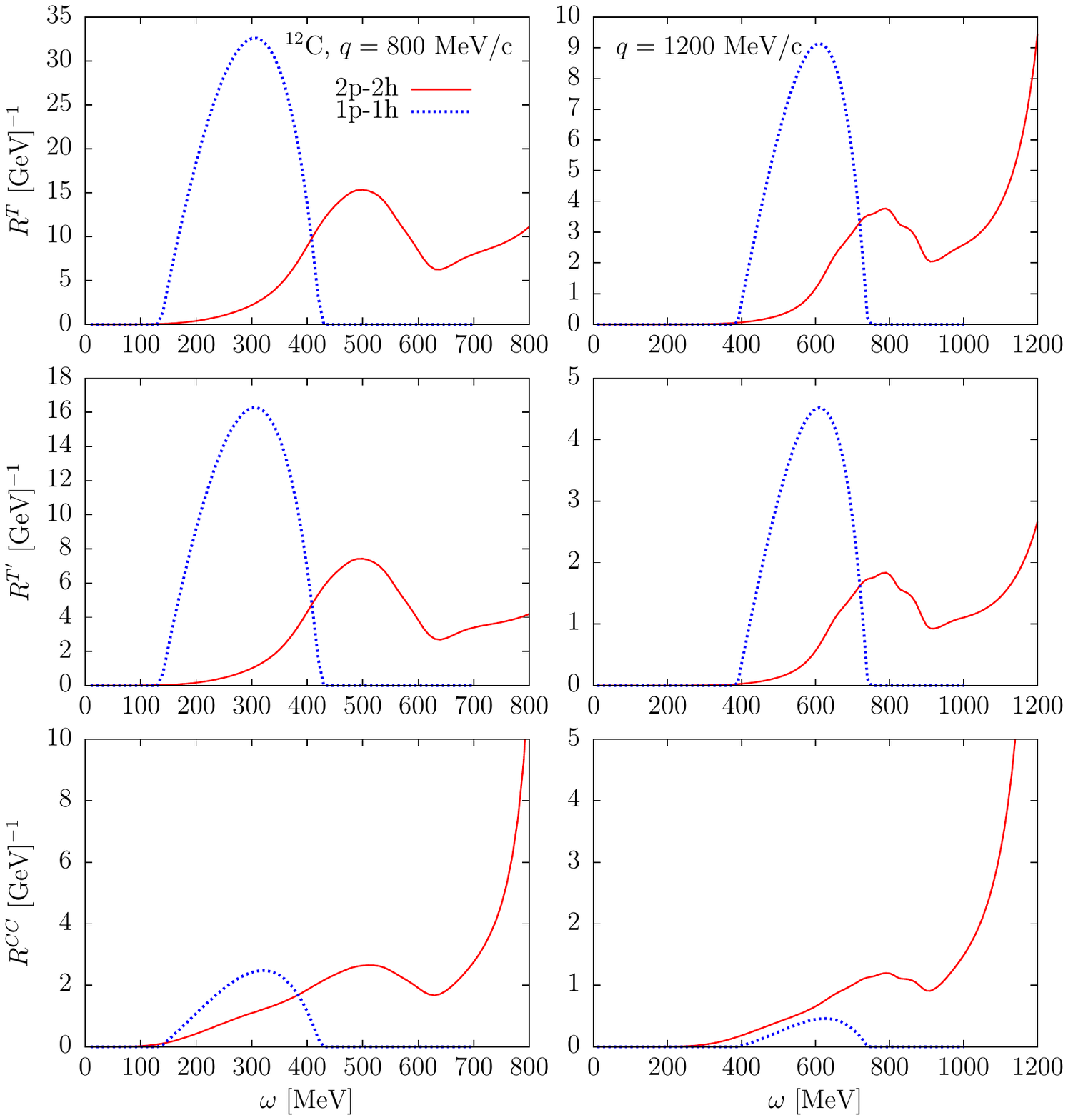}
   \vspace{-4.cm}
\caption[]{Comparison between 1p1h and 2p2h response functions for CC neutrino scattering off $^{12}$C for two values of the
momentum transfer \cite{Simo:2016ikv}.}
\label{fig1}
    \end{figure}
In Fig. \ref{fig1} we show the 1p1h and 2p2h responses as functions of the energy transfer $\omega$ for $q$ = 800 and 1200 MeV/c. The 1p1h responses are computed in the RFG and only contain
the one-body current. For these values of $q$ the MEC effects in the $T$ and $T'$ channels are large, the 2p2h strength at the maximum of the peak being around 1/2 of the 1p1h response.
Moreover, the MEC effects are similar in the $T$ and $T'$ responses.
The $CC$ response appears to be extremely sensitive to MEC effects, which are even larger than the one-body response. 
However, the contribution of this response to the cross section and the ones associated to the $CL$ and $LL$ channels largely cancel out, so that the net charge/longitudinal cross section is generally smaller than the transverse ones.
The balance of the different response functions of course
depends on the kinematics.
\begin{figure}[h]
\includegraphics[scale=0.5]{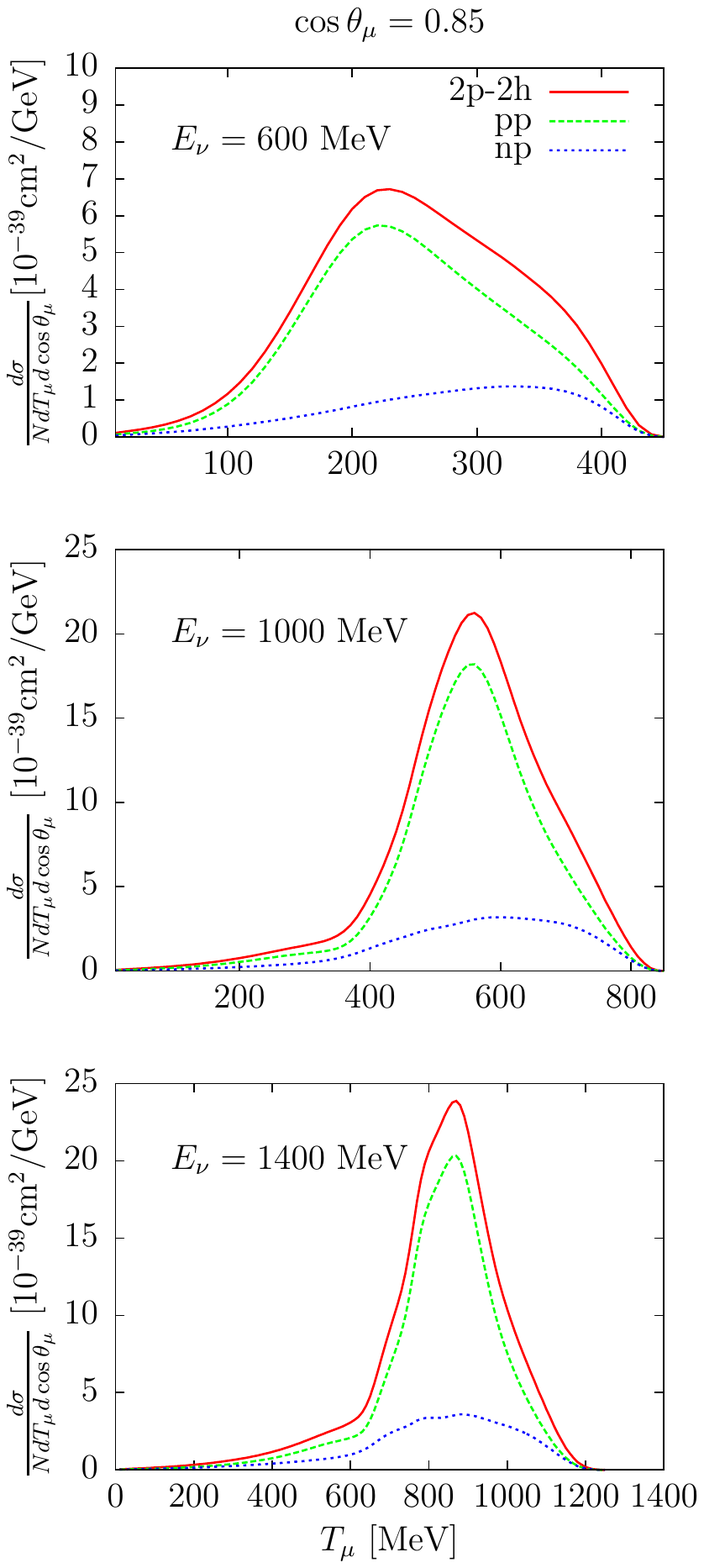}
   \vspace{-4.cm}
\caption{ Doubly differential 2p2h neutrino cross section per neutron of
  $^{12}$C, for fixed muon scattering angle and for three neutrino
  energies, as a function of the muon kinetic energy.  The separate
  $np$ and $pp$ channels are shown \cite{Simo:2016ikv}. }
\label{fig7}
\end{figure}

The separate $pp$ and $np$ emission channels in the differential neutrino cross
 section are shown in fig. \ref{fig7} for three different values of the neutrino energy. Proton-proton final states clearly dominate the 2p2h cross section. The $pp/np$ ratio is
 around 5-6 near the maximum, but its precise value depends on the
 kinematics. Note that the $np$ distribution is shifted towards higher
 muon energies respect to the $pp$ one.
\begin{figure}[ht]
\begin{center}
\includegraphics[scale=0.2, angle=0]{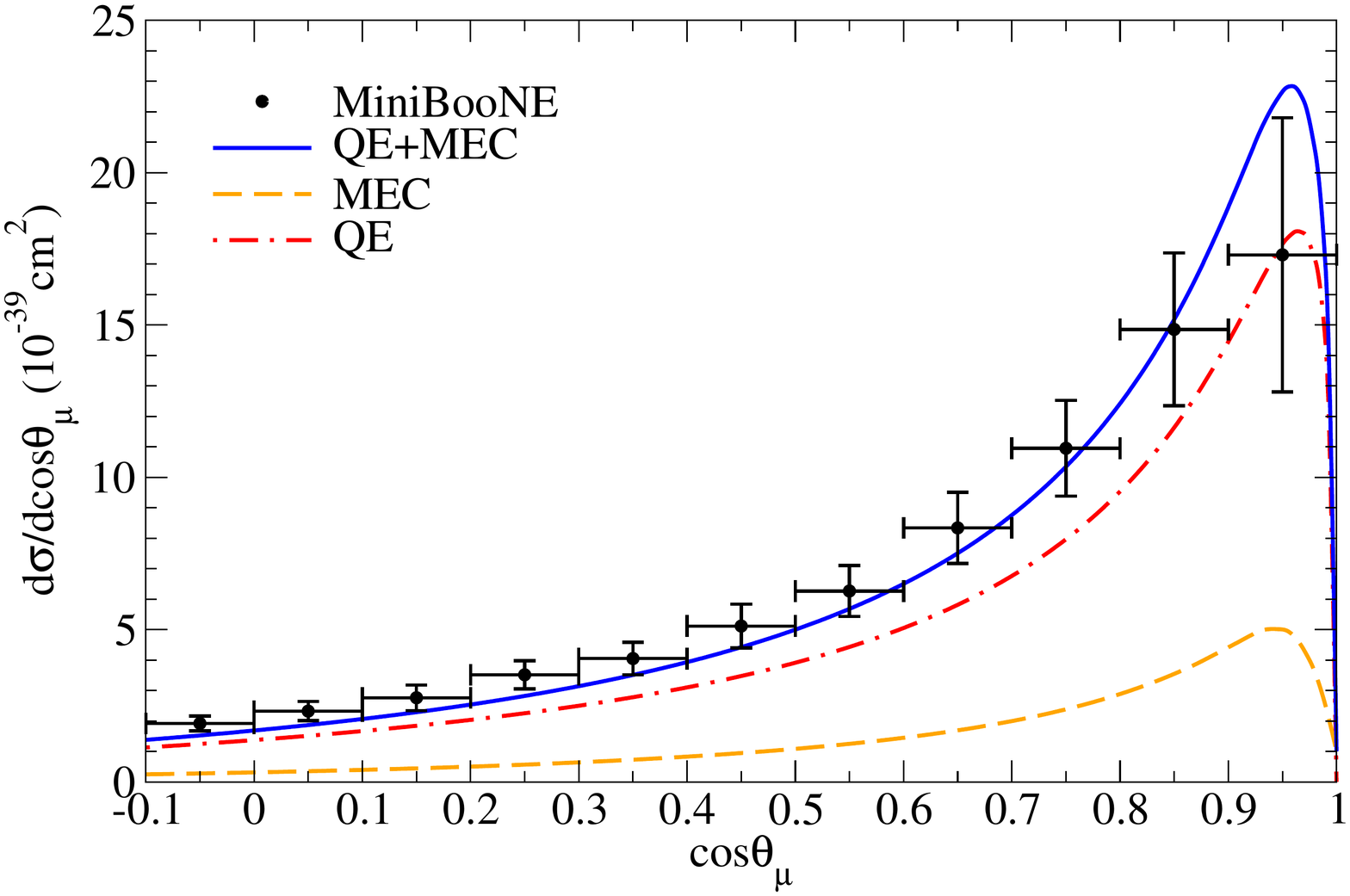}\hspace*{-0.05cm}\includegraphics[scale=0.2, angle=0]{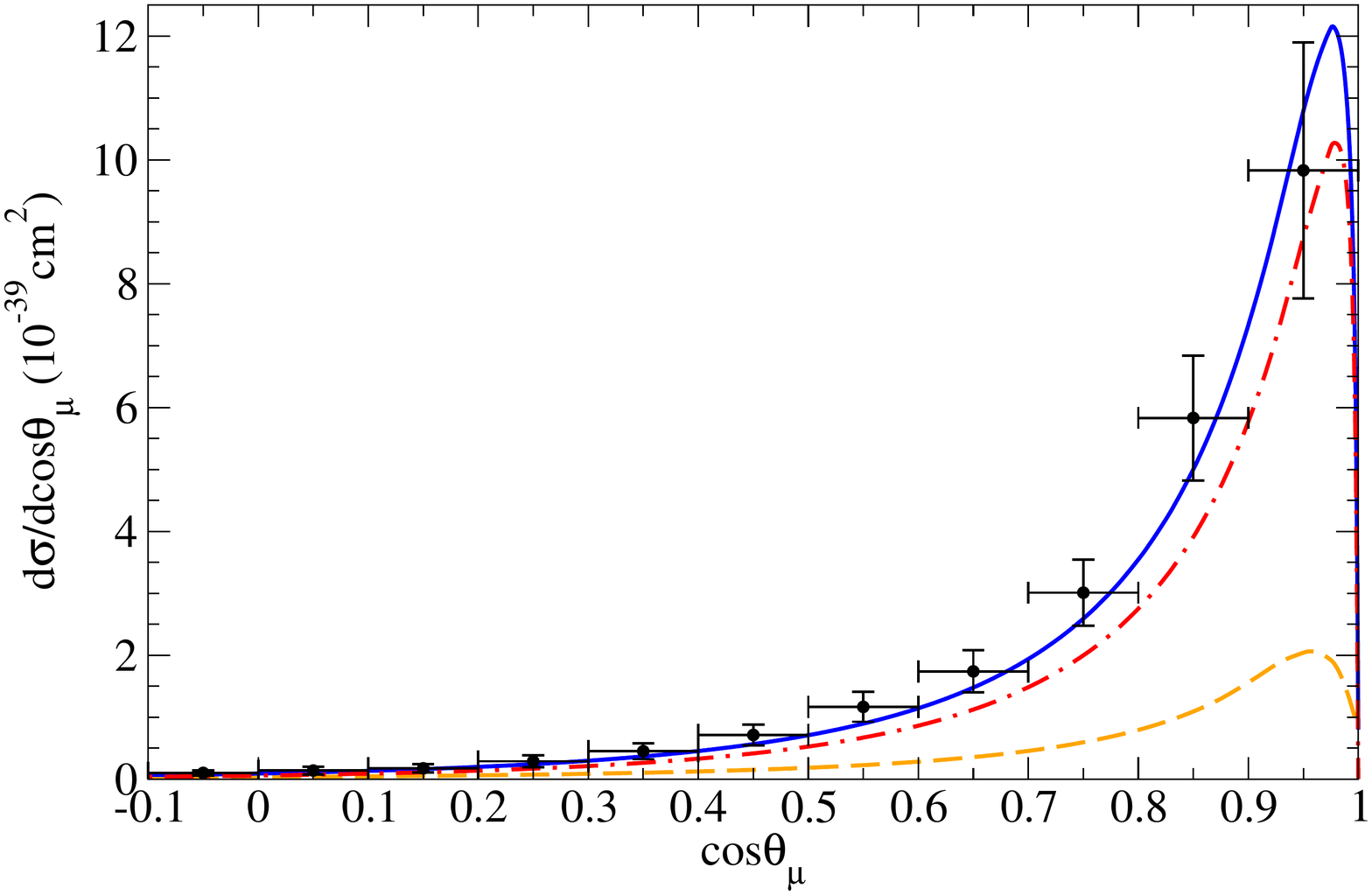}\\
\includegraphics[scale=0.2, angle=0]{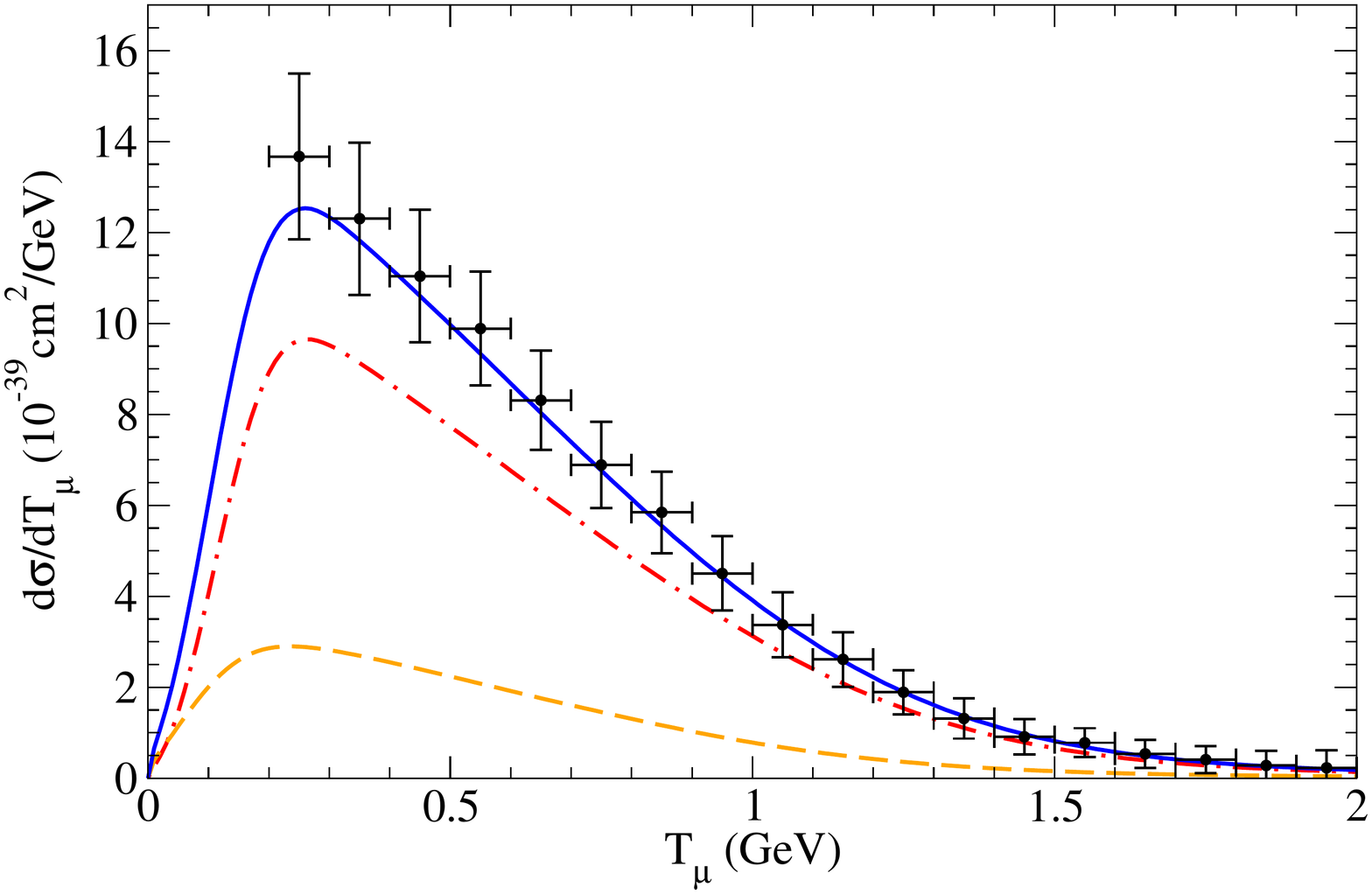}\hspace*{-0.05cm}\includegraphics[scale=0.2, angle=0]{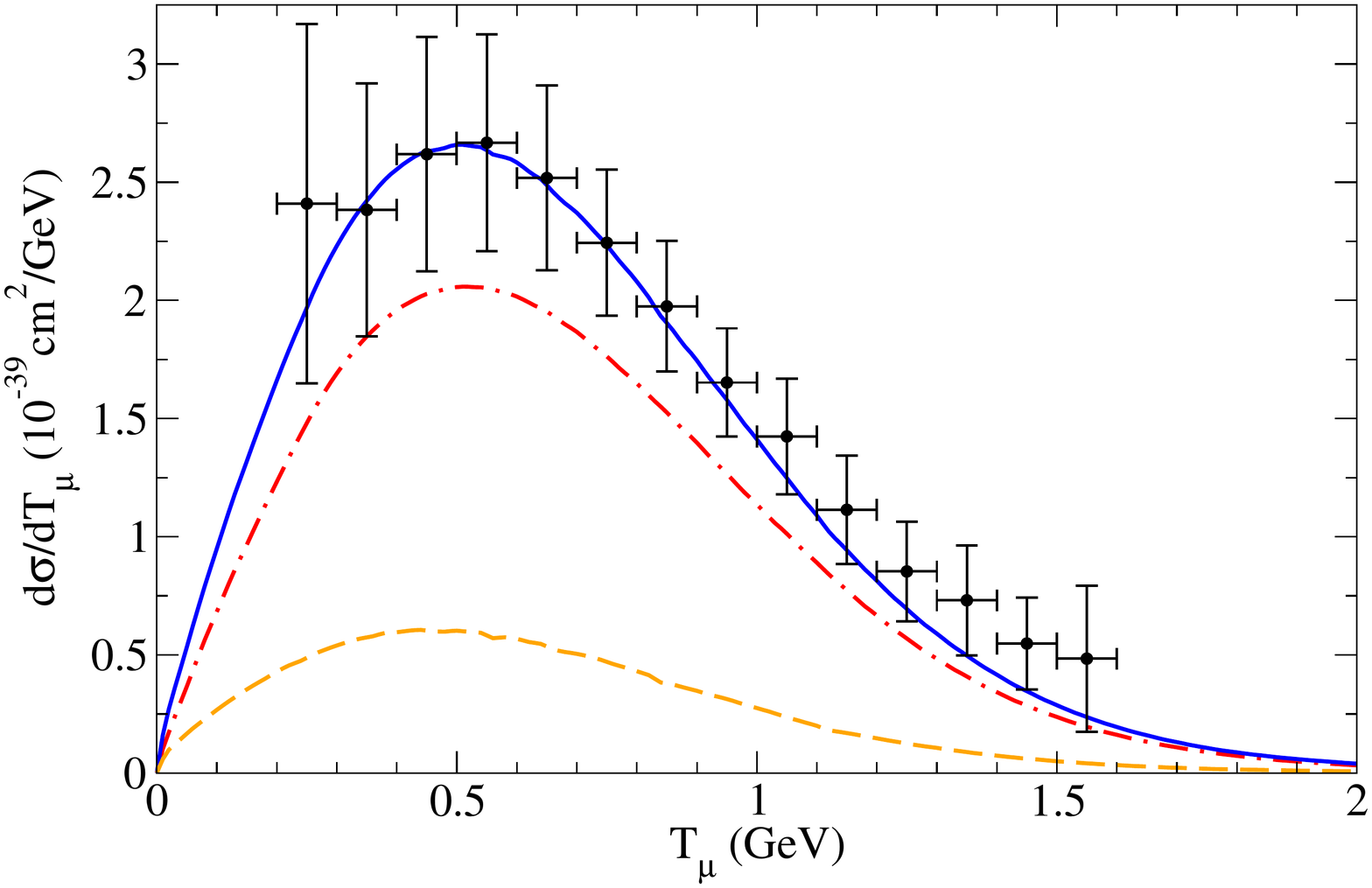}
\begin{center}
\vspace{-1cm}
\end{center}
\end{center}
\caption{MiniBooNE flux-averaged CCQE $\nu_\mu$-$^{12}$C ($\bar\nu_\mu$-$^{12}$C) differential cross section per nucleon as a function of the muon scattering angle (top panels) and of
  the muon kinetic energy (bottom panels). The left panels correspond to neutrino cross sections and the right ones to antineutrino reactions.
  Data are from \cite{AguilarArevalo:2010zc,AguilarArevalo:2013hm}.}
\label{CS_single}
\end{figure}

We now present some comparison with experimental data. In order to do that, we combine the present calculation of the 2p2h MEC contribution with the SuSAv2 model, used to describe the QE and $\Delta$-excitation regions. We refer to this as ``SuSAv2-MEC'' model.
Due to the limited  space we choose some representative examples. The interested reader can find more results in Ref. \cite{Megias:2016fjk}.

In Fig.~\ref{CS_single} the MiniBooNE flux-averaged CCQE $\nu_\mu (\overline{\nu}_\mu)-^{12}$C differential cross section per nucleon is shown as a function of the muon scattering angle (top panels) and the muon kinetic energy (bottom panels). Panels on the left (right) correspond to neutrinos (antineutrinos). The mean beam energy in the MiniBooNE experiment is 0.788 GeV in the neutrino mode and 0.665 GeV in the antineutrino mode.
As shown, the present model provides an excellent representation of the experimental cross section. In all of the cases the MEC contribution is essential in order to reproduce the data and it amounts to about 20-25\% of the total response for neutrino, 15-20\% for antineutrino. 
Analogous results~\cite{Megias:2016fjk}, not shown here, are found for the CCQE cross section measured in the T2K experiment, where the neutrino energy is close to the MiniBooNE one, but the flux is less broadly distributed.

\begin{figure}[h]
\begin{center}
\includegraphics[scale=0.2, angle=270]{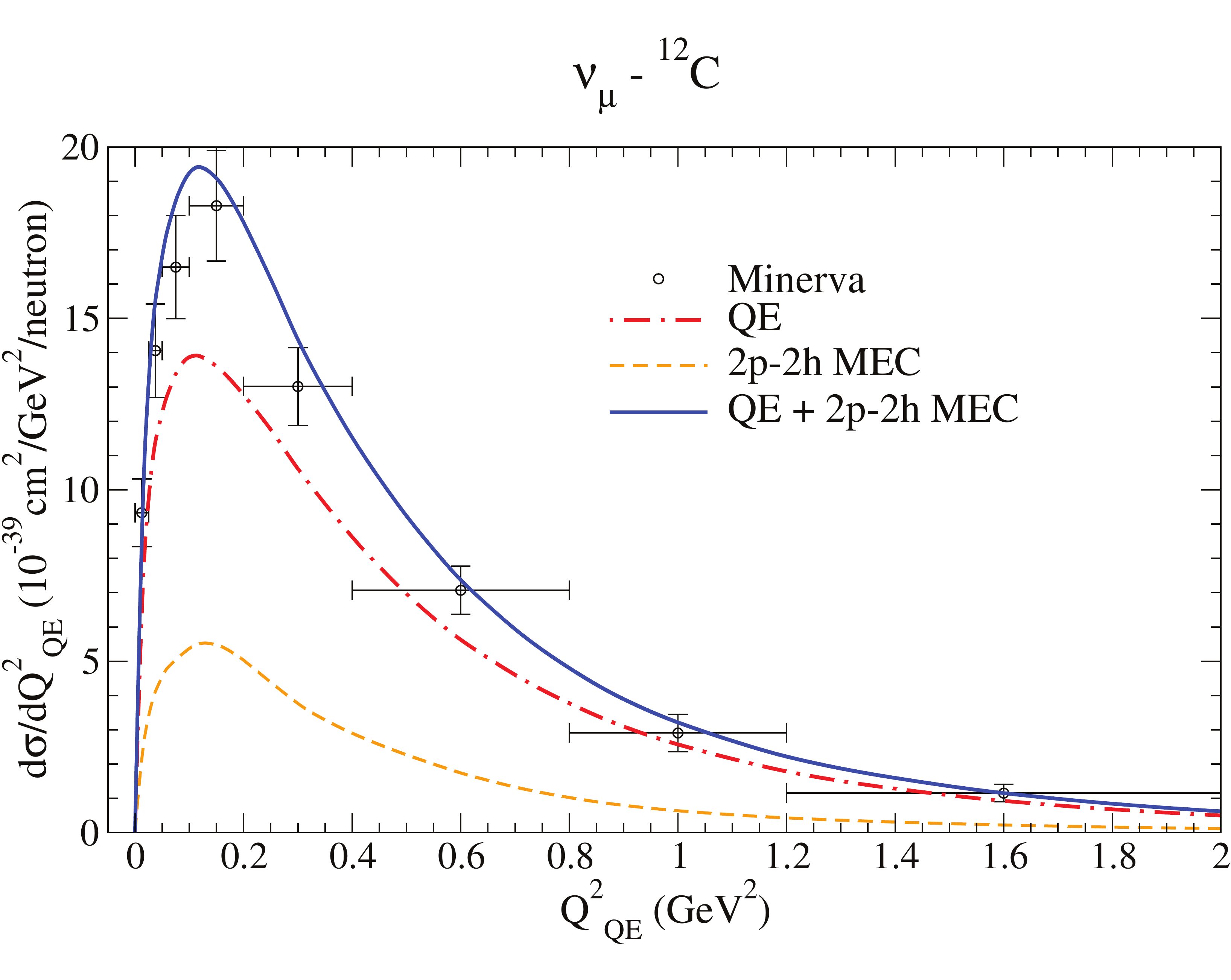}\\
\includegraphics[scale=0.2, angle=270]{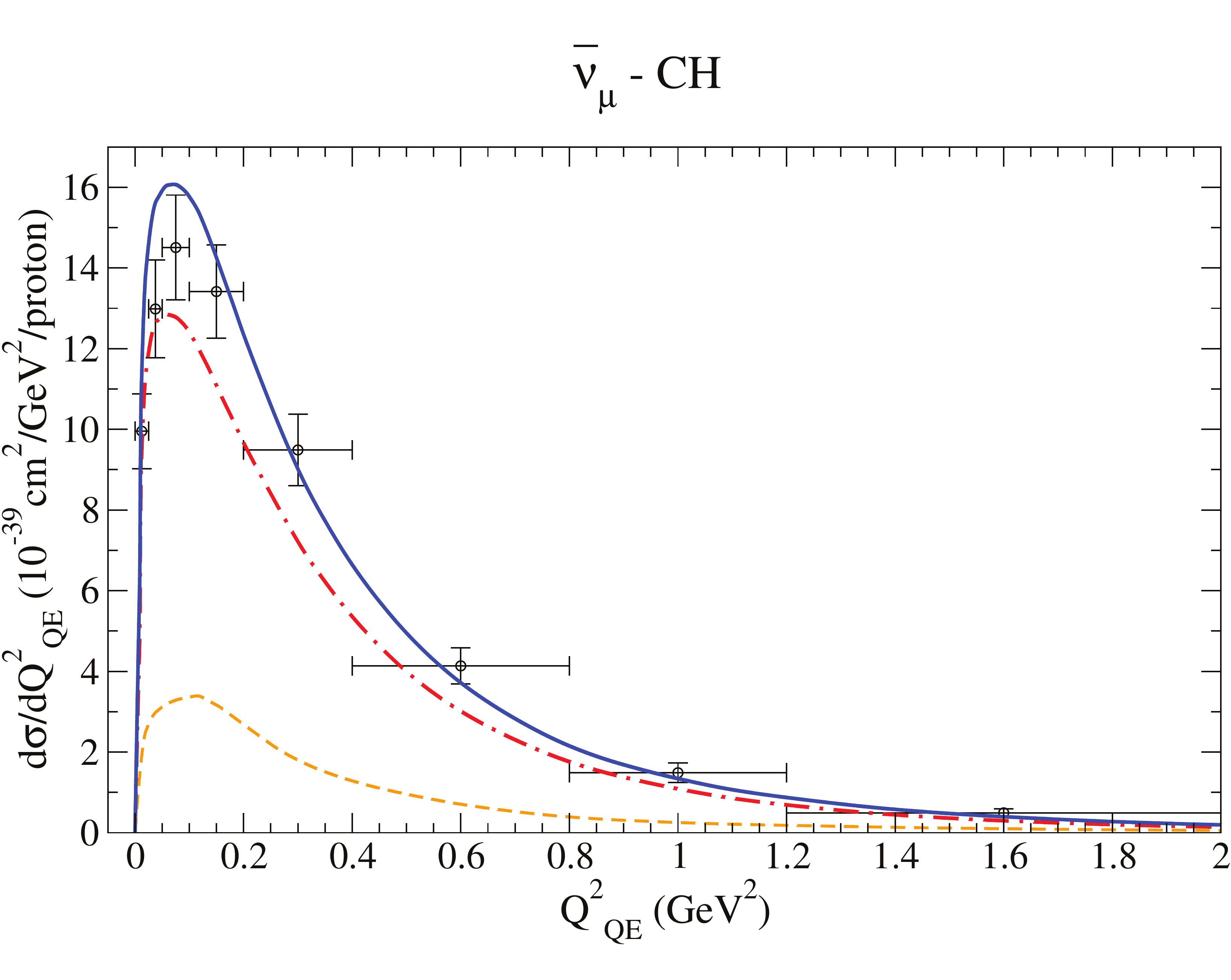}
\vspace{1cm}
\end{center}
\caption{Flux-folded $\nu_\mu-^{12}$C CCQE (upper panel) and $\bar\nu_\mu-$CH (lower panel) scattering cross section per
target  nucleon  as  a  function  of $Q^2_{QE}$ and  evaluated  in  the
SuSAv2  and  SuSAv2+MEC  models.    MINERvA  data  are from~\cite{MINERVA,MINERVApriv}.}
\label{Minerva_numu}
\end{figure}
A similar trend is shown in Fig.~\ref{Minerva_numu}, where the MINERvA flux-averaged CCQE $\nu_\mu (\overline{\nu}_\mu)$ differential cross section per nucleon is displayed as a function of the reconstructed four-momentum $Q^2_{QE}$. The top panel refers to $\nu_\mu-^{12}$C whereas the bottom panel contains predictions and data for $\overline{\nu}_\mu-$CH~\footnote{These data correspond to a new analysis of the MINERvA collaboration~\cite{MINERVApriv}, based on the updated NuMI flux prediction~\cite{Aliaga:2016oaz}, and exceed by $\sim 20\%$ the ones published in~\cite{MINERVA}.}. 
Note that the mean energy of the MINERvA flux is much higher than the MiniBooNE one, about 3.5 GeV for both $\nu_\mu$ and $\overline{\nu}_\mu$. Also in this case the contribution of the 2p2h MEC is needed in order to reproduce the experimental data.

\begin{figure}[ht]
\begin{center}
\includegraphics[scale=0.25, angle=0]{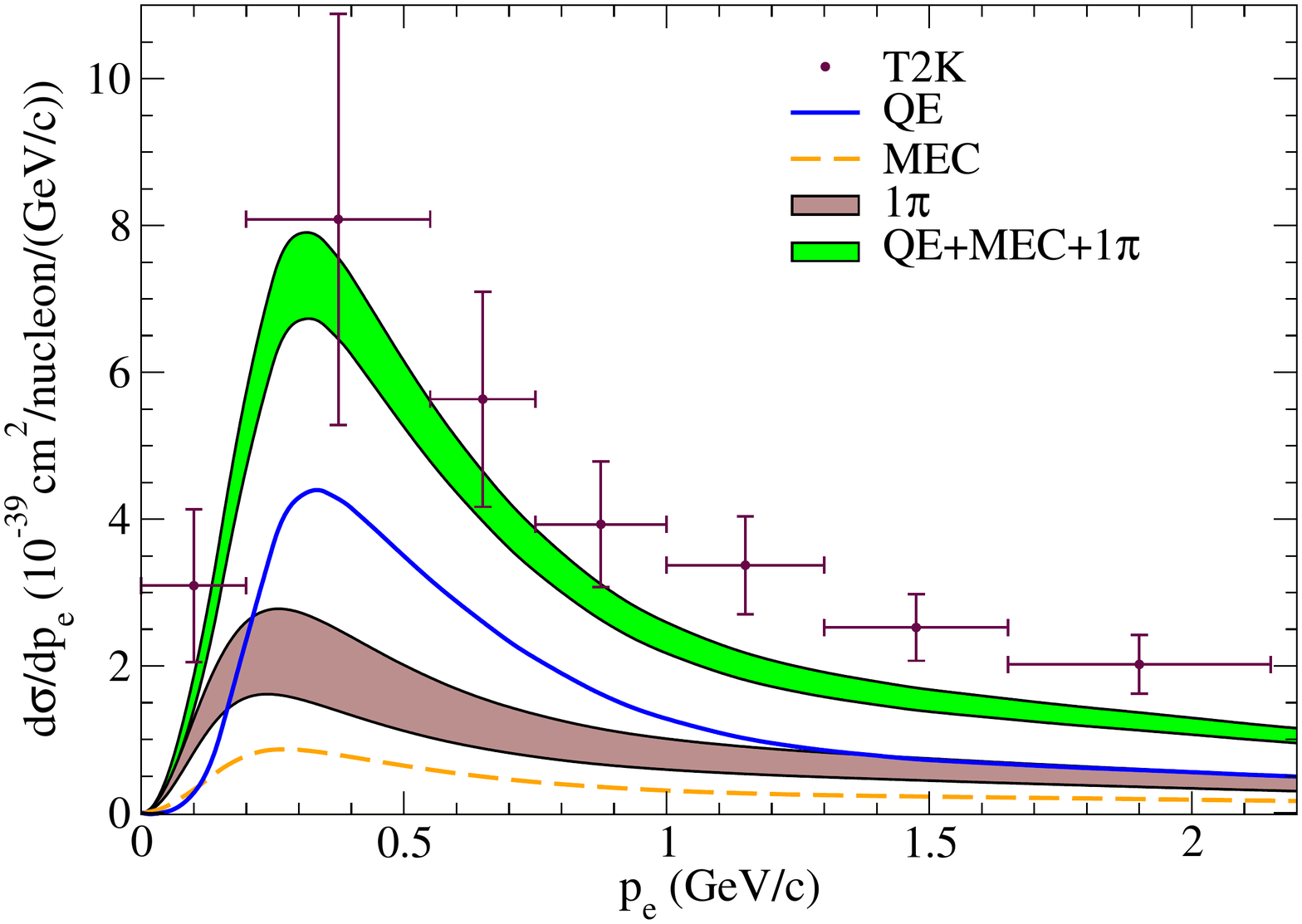}
\begin{center}
\vspace{-1cm}
\end{center}
\end{center}
\caption{The CC-inclusive T2K flux-folded $\nu_e-^{12}$C differential cross section per nucleon evaluated in the SuSAv2+MEC model is displayed as a function of the electron momentum. The separate contributions of the QE, 1$\pi$
  and 2p2h MEC are displayed \cite{Ivanov:2015aya}. The data are from~\cite{T2Kinclelectron}.}
\label{inclusive_T2K_e}
\end{figure}

In Fig.~\ref{inclusive_T2K_e} the inclusive electron-neutrino flux-averaged single differential cross section on $^{12}$C measured by the T2K experiment is shown as a function of the electron momentum. The neutrino energy is very similar to the of the MiniBooNE experiment. However, here the data are inclusive, namely only the final electron is observed and pions, or any other product of the reaction, can be present in the final state.
We show the separate contributions corresponding to the QE response, the 2p2h MEC, pionic and the total response. The colored bands correspond to the estimated uncertainty in the scaling analysis extended to the $\Delta$ region (see \cite{Ivanov:2015aya}).
Although the role associated with the $\Delta$ resonance is essential, the model is underestimating the data, especially for increasing values of the electron momentum.
This indicates that other inelastic channels, not taken into account in the present description, may play a significant role in explaining these data.
Work along this line is presently in progress.

\section{Conclusions}
\label{sec:conclusions}

Two-particle-two-hole excitations induced by meson-exchange currents play a very important role in interpreting neutrino scattering data. We have illustrated this point by comparing the SuSAv2+MEC predictions with data on neutrino and antineutrino scattering off $^{12}$C corresponding to three different experiments: MiniBooNE, MINERvA and T2K.

We have also studied the separate charge channels contributing to this process and shown that for neutrino scattering $pp$ final states give a contribution five to six times larger than the $np$ ones. Having the separate isospin contributions will allow us to apply this formalism to asymmetric nuclei $N\ne Z$,  of interest for neutrino experiments based on $^{40}$Ar.

Future developments will focus on the extension of the model to different nuclei and to the inelastic region, of paramount importance for future experiments.

\section*{Appendix}
In this Appendix we provide the explicit expressions of some quantities used in the text.

The factor $\sigma_0$ appearing in Eq.(\ref{sigma}) is
\begin{equation}
\sigma_0=
\frac{G_F^2\cos^2\theta_c}{2\pi^2}
\left(k^\prime \cos\frac{\tilde\theta}{2}\right)^2 ,
\end{equation}
where $G_F$ is the Fermi
constant, $\theta_c$ the Cabibbo angle and $\tilde\theta$ the generalized scattering angle, defined as
\begin{equation}
\tan^2 \frac{\tilde\theta}{2} = \frac{|Q^2|}{(\epsilon+\epsilon')^2-q^2} .
\end{equation}

The kinematical factors in Eq.(\ref{Spm}) are given by
\begin{eqnarray}
V_{CC}
&=&
1-\delta^2\tan^2\frac{\tilde{\theta}}{2}
\label{vcc}\\
V_{CL}
&=&
\frac{\omega}{q}+\frac{\delta^2}{\rho'}\tan^2\frac{\tilde{\theta}}{2}
\\
V_{LL}
&=&
\frac{\omega^2}{q^2}+
\left(1+\frac{2\omega}{q\rho'}+\rho\delta^2\right)\delta^2
\tan^2\frac{\tilde{\theta}}{2}
\\
V_{T}
&=&
\tan^2\frac{\tilde{\theta}}{2}+\frac{\rho}{2}-
\frac{\delta^2}{\rho'}
\left(\frac{\omega}{q}+\frac12\rho\rho'\delta^2\right)
\tan^2\frac{\tilde{\theta}}{2}
\\
V_{T'}
&=&
\frac{1}{\rho'}
\left(1-\frac{\omega\rho'}{q}\delta^2\right)
\tan^2\frac{\tilde{\theta}}{2}\ .
\label{vtp}
\end{eqnarray}
In Eqs.~(\ref{vcc}--\ref{vtp}) we have defined
\begin{eqnarray}
\delta &=& \frac{m_l}{\sqrt{|Q^2|}}\\
\rho &=& \frac{|Q^2|}{q^2}\\
\rho' &=& \frac{q}{\epsilon+\epsilon'}\ ,
\end{eqnarray}
where $m_l$ is the final charged lepton mass.


\end{document}